\documentclass[12pt]{article}
\usepackage{rotating}
\usepackage{indentfirst}
\usepackage{cite}
\usepackage{doublespace}
\usepackage{citesupernumber}
\usepackage{epsfig}
\title{Markov Chains of Infinite Order and Asymptotic Satisfaction of
Balance: Application to the Adaptive Integration Method}

\author{David J.\ Earl and Michael W.\ Deem\\
Departments of Bioengineering and Physics \& Astronomy\\
Rice University\\
6100 Main Street---MS 142\\
Houston, TX\ \ 77005--1892
}

\begin{document}
\maketitle
Corresponding author: M.\ W.\ Deem, mwdeem@rice.edu, fax: 713--348--5811.
\clearpage
\newpage

\begin{abstract}
Adaptive Monte Carlo methods can be viewed as implementations
of Markov chains with infinite memory.  We derive a general
condition for the convergence of a Monte Carlo method whose history 
dependence is contained within the simulated 
density distribution.  In convergent cases,
our result implies that the balance condition need only be
satisfied asymptotically.  As an example, we show that 
the adaptive integration method converges.
\end{abstract}

\section{Introduction}
\label{sec:introd}

Adaptive Monte Carlo methods that change the sampling
strategy based upon statistics collected 
on the fly have been shown to be very powerful
in a number of interesting
applications.\cite{swendsen6,swendsen7,swendsen8,swendsen9,swendsen10}
Typically these adaptive methods use the 
statistics collected during a run
to construct an importance sampling potential
that is intended to remove the most significant barriers to sampling
in the problem.
These methods, however, have been criticized
by various authors due to their lack of satisfying detailed
balance.  
Although the use of adaptive simulation
methods is growing, and their success has been demonstrated 
in a number of cases,\cite{swendsen9}
widespread acceptance of the correctness of the approach
within the simulation community has been hampered by these
questions of detailed balance and so of convergence of the simulated
distribution in phase space.
In most of these methods, it is clear that there is
at least one fixed point of the distribution, the Boltzmann,
possibly modified by an importance sampling factor that is
itself a functional of the Boltzmann distribution.  As the Monte
Carlo algorithm will be started from an arbitrary initial condition,
it is of interest to know whether these algorithms will converge
to the Boltzmann, or any other, fixed  point.  Such
fixed point analysis has not been performed for this class of algorithms
and is the subject of this paper.

As an archetypal example of these adaptive Monte Carlo methods, we consider
the adaptive integration scheme of 
Fasnacht, Swendsen, and Rosenberg.\cite{swendsen}  
In this method, we focus on one order parameter, $\lambda$, of the
system that leads to the most significant
barriers. We construct an
estimate of the probability distribution of $\lambda$, and we
use the inverse of this probability
as the importance sampling function:
\begin{equation}
P(\lambda_0) = 
\langle \delta(\lambda({\bf x}) - \lambda_0) \rangle
=
\frac{
\int d{\bf x} e^{-\beta U({\bf x})}
    \delta(\lambda({\bf x}) - \lambda_0)
}
{
\int d{\bf x} e^{-\beta U({\bf x})}
}  \ ,
\label{eq0}
\end{equation}
where $\beta$ is the inverse temperature, and ${\bf x}$ is
a vector in $ 3 n$ dimensions.  We define
\begin{equation}
e^{-\beta F(\lambda_0)} = \int d{\bf x} e^{-\beta U({\bf x})}
    \delta(\lambda({\bf x}) - \lambda_0) \ .
\end{equation}
Then the probability distribution is given by
\begin{equation}
P(\lambda_0) = \frac{e^{-\beta F(\lambda_0)}}{  Z}  \ ,
\end{equation}
where the configurational integral is 
$Z = \int d{\bf x} e^{-\beta U({\bf x})}$.
Now, consider 
\begin{eqnarray}
\frac {d \ln P(\lambda_0)}{d \lambda_0} &=&
\frac{
\int d{\bf x} e^{-\beta U({\bf x})}
    \frac{d}{d \lambda_0} \delta(\lambda({\bf x}) - \lambda_0)
}
{
\int d{\bf x} e^{-\beta U({\bf x})}
     \delta(\lambda({\bf x}) - \lambda_0)
}
\nonumber \\
&=& 
\frac{
\int d{\bf x} 
\delta(\lambda({\bf x}) - \lambda_0)
\frac{d}{d \lambda_0} 
e^{-\beta U({\bf x})}
}
{
\int d{\bf x} e^{-\beta U({\bf x})}
     \delta(\lambda({\bf x}) - \lambda_0)
}\nonumber \\
&=& 
-\beta \frac{
\int d{\bf x} 
e^{-\beta U({\bf x})}
\frac{d U}{d \lambda_0} 
\delta(\lambda({\bf x}) - \lambda_0)
}
{
\int d{\bf x} e^{-\beta U({\bf x})}
     \delta(\lambda({\bf x}) - \lambda_0)
}\nonumber \\
&=& -\beta \left\langle \frac{d U}{d \lambda_0} \right\rangle_{\lambda_0} \ ,
\end{eqnarray}
where we have used the fact that both $\lambda$ and $U$ are 
functions of ${\bf x}$, and $d U / d \lambda$ really  means
$\nabla U \cdot d {\bf x} / d \lambda$.
Thus, we come to the thermodynamic integration formula
for the free energy as a function of $\lambda$:
\begin{eqnarray}
F(\lambda_0) &=& \int d\lambda \left\langle \frac{d U }{d \lambda}
 \right\rangle
\nonumber \\ 
&=& \int_{\lambda_{\min}}^{\lambda_0} d\lambda'
\frac{
\int d{\bf x} e^{-\beta U({\bf x})} \frac{d U}{d \lambda}
    \delta(\lambda({\bf x}) - \lambda')
}
{
\int d{\bf x} e^{-\beta U({\bf x})}
    \delta(\lambda({\bf x}) - \lambda')
} \ .
\label{eq1}
\end{eqnarray}
We desire the observed distribution in the Monte Carlo scheme
to be 
\begin{equation}
\rho({\rm \bf x}) = (\rm const) e^{-\beta U({\bf x })} /
 P(\lambda({\bf x }))
= ({\rm const} ) e^{-\beta U({\bf x }) + \beta F({\bf x })} \ .
\label{eq2}
\end{equation}
As the simulation is performed, the estimated value of importance
sampling free energy, $\hat F(\lambda)$, is constructed
from Eqn.\ \ref{eq2} and used in Eqn.\ \ref{eq1}.  For example,
if the distribution to date is $\rho({\bf x})$, then we have
\begin{equation}
\hat F(\lambda_0)
= \int_{\lambda_{\min}}^{\lambda_0} d\lambda'
\frac{
\int d{\bf x} \rho({\bf x}) \frac{d U}{d \lambda}
    \delta(\lambda({\bf x}) - \lambda')
}
{
\int d{\bf x} \rho({\bf x})
    \delta(\lambda({\bf x}) - \lambda')
} \ .
\label{eq3}
\end{equation}
  In the
context of a Metropolis algorithm, the acceptance criterion
would be
\begin{equation}
{\rm acc}(o \to n)
 = \min\left\{1, e^{-\beta U({\bf x}^n)
+\beta U({\bf x}^o) 
+ \beta \hat F(\lambda({\bf x}^n)) 
- \beta \hat F(\lambda({\bf x}^o))
}
 \right\} \ .
\label{eq4}
\end{equation}
Although we have written transition probabilities that satisfy
detailed balance in Eq.\ \ref{eq4},
our analysis is equally applicable to
transition probabilities that satisfy only balance.\cite{Deem_balance}
Since the importance sampling function is changing with step number,
due to the updates to density and so to $\hat F$, this 
adaptive algorithm does not
satisfy detailed balance.  It is clear, however, that if
the density has converged to the Boltzmann, 
$\rho({\bf x}) = (\rm const \mit) e^{-\beta U({\bf x}) + \beta F({\bf x})}$,
then the estimation in Eqn.\ \ref{eq3} is exact, and
the acceptance criterion in Eqn.\ \ref{eq4} is exact
and constant for future steps.
Thus, the desired exact answer, Eqn.\ \ref{eq2}, is a fixed point
of this algorithm.  We also note that if the observed
density distribution is not exact, but if
$\rho({\bf x}) / \rho({\bf y}) = e^{-\beta U({\bf x})
+ \beta U({\bf y})}$ whenever $\lambda({\bf x}) = \lambda({\bf y})$, 
then the estimated importance sampling function, Eqn.\ \ref{eq3},
is also exact.  This property will prove to be useful.

The Markov process underlying a Monte Carlo algorithm with
acceptance probabilities such as Eqn.\ \ref{eq4} has memory,
because the process depends on all past states through the observed
density distribution $\rho({\bf x})$.  Technically, this
is a Markov chain of infinite order.\cite{Harris}
Markov chains of infinite order have a solid body of
convergence results when the dependence on the past,
essentially, decays exponentially fast (technically, when
they are continuous with respect to the past).\cite{Fernandez1}
The Markov process in adaptive integration is dramatically
discontinuous with respect to the past: the first point observed
is just as important as the most recent point observed in the
measured density distribution.

We here consider the infinite order
Markov process that occurs in adaptive Monte Carlo.
In Sec.\ \ref{sec:theory}, we derive a general 
condition on the transition matrices for adaptive Monte Carlo
that guaranties convergence to a unique distribution.
Although we often use continuous notation, technically
we limit discussion to finite, discrete spaces, in the belief that
the continuous limit exists as the grid spacing goes to zero.
In Sec.\ \ref{sec:application}, we examine the special case
of adaptive integration Monte Carlo, showing convergence to a
unique distribution occurs.  We discuss our findings
in Sec.\ \ref{sec:discussion}.

\section{Theory}
\label{sec:theory}
We wish to find the conditions under
which a Markov chain of infinite order
converges to a unique limiting distribution.
We consider the chain to possess a 
transition probability that depends on the current and future
state and on the density constructed from all previously observed states,
as in Eqns.\ \ref{eq3}--\ref{eq4}.
We assume that the transition probabilities satisfy a generalization
of the irreducibility condition: we assume there is $c>0$ such that
\begin{equation}
{\rm Prob}[{\bf x}(t), \rho(t) \to {\bf y}(t+m)] > c 
\label{irr}
\end{equation}
for all ${\bf x}$, ${\bf y}$, $\rho$, and $t$ for some fixed $m$.
This is a precise statement of our desire that the process
be ergodic, with mixing time $m$.
Thus,  we have ${\rm Prob}[{\bf x}(t)] > c$ for all times $t>m$, because we
can apply Eqn.\ \ref{irr} to each of the initial states $1, \ldots, m-1$,
and then iterate to conclude 
${\rm Prob}[{\bf x} (m'+m)]
= \sum_{{\bf x}'}
{\rm Prob}[{\bf x}' (m'), \rho(m') \to {\bf x}(m+m')] {\rm Prob}[{\bf x}'(m')]
\ge c  \sum_{{\bf x}'} {\rm Prob}[{\bf x}'(m')] = c$.
In fact, we consider a larger value of $m$, so that
any given Markov chain has an observed density that equals the
expected probability distribution to $O(1/\sqrt m)$.\cite{clt}
Now consider a Markov chain that has run for $M$ Monte Carlo steps,
$M \gg m$.
For this process, it will be true that
\begin{equation}
\rho (M+m) = \rho (M) + O(m/M)
\cong \rho (M)
  \ .
\end{equation}
Thus, the transition matrix $A[\rho(M)]$ will be roughly 
constant during this time interval, since the density
itself is not changing much, and assuming the transition matrix
depends smoothly on the density.   The distribution of
new states during this time period will converge to
\begin{eqnarray}
{\rm Prob}[{\bf x}(M+m) ] & = &
A[\rho(M+m-1)]
A[\rho(M+m-2)] \ldots
A[\rho(M)] \rho(M)
\nonumber \\ 
& =&
A[\rho(M)]^m \rho(M) + O(m^2/M)
\nonumber \\ 
& = &
\rho^{*} [\rho (M)] + O (r^m, m^2/M) \ .
\end{eqnarray}
Here $r < 1 $ is the second largest generalized eigenvalue of 
$A[\rho(M)]$.\cite{Grantmacher}
The probability distribution
is driven to the limiting distribution of the transition
matrix for large $m$:
${\rm Prob}[{\bf x}(M+m)] \rightarrow \rho ^{*} [\rho (M)]$. 
By the Frobenius-Perron theorem,\cite{Grantmacher} this
limiting distribution depends on the measured density,
but not on the state at $M$:
$\rho^* = 
\lim_{m \to \infty} A^m[\rho(M)]\rho(M) =
\lim_{m \to \infty} A^m[\rho(M)]\rho$ for any $\rho$.
By the central limit theorem of Markov processes,\cite{clt}
any likely instance of this probability distribution will 
be accurate to $O(1/\sqrt m)$.
Thus, the contribution of these $m$ steps to the history-dependent
density is
\begin{equation}
\rho (M+m) = \frac{M}{M+m} \rho (M) + \frac{m}{M+m} \rho ^ {*} [\rho (M)]
 + \frac{m}{M+m} O(\frac{m^2}{M}, r^m, 1/\sqrt m) \ ,
\label{eq7}
\end{equation}
Since we consider the limit of $1 \ll m^2 \ll M$, we may drop the
error terms.  We let $u= \frac{m}{M+m}$.
Then by the contraction mapping theorem on compact spaces\cite{cmt}
there will be
a fixed point to Eqn.\ \ref{eq7} if there is a metric, $D$, such
that
\begin{equation}
D\left[ (1-u) \rho_1  + u \rho ^{*} [\rho_1 ],
(1-u)\rho_2  + u \rho ^{*} [ \rho_2 ]
\right]
\label{aldous}
\end{equation}
is initially decreasing as $u$ increases from 0,
for any \emph{arbitrary} $\rho_1$ and $\rho_2$.
If this condition is satisfied, the fixed point
exists and is unique for our finite, discrete system.\cite{cmt}
We note that if the following is satisfied for
arbitrary $\rho_1$ and $\rho_2$, then
Eqn.\ \ref{aldous} is automatically satisfied for small $u$:
\begin{equation}
D\left[ \rho ^{*} [\rho_1] , \rho ^{*} [\rho_2 ] \right]
< D [\rho_1 , \rho_2]
\end{equation}
Alternatively, we can consider the uniqueness and existence of
the mapping $\rho_{n+1} = \rho^*(\rho_n) $ for 
arbitrary $\rho_0$.

\section{Application to Adaptive Monte Carlo}
\label{sec:application}

The general condition, Eqn.\ \ref{aldous}, seems difficult to check
for an arbitrary functional dependence on the measured density, 
$\rho^*[\rho]$.
We, thus, specialize consideration to the adaptive integration 
method.  We rewrite Eqn.\ \ref{eq7} as
\begin{equation}
\rho (t+\Delta t) = (1 - \frac{\Delta t}{t}) \rho + \frac{\Delta t}{t} \rho ^{*} [\rho (t)]
\end{equation}
where $\Delta t = m$.  Assuming that this difference equation is
well-approximated by a differential equation, we find
\begin{equation}
\frac{d \rho}{d t} \Delta t = - \frac{\Delta t \rho}{t} + \frac{\Delta t}{t} \rho^{*}[\rho]
\end{equation}
\begin{equation}
\frac{d \rho}{d t} = \frac{1}{t} ( \rho^{*}[\rho] - \rho)
\end{equation}
\begin{equation}
\frac{d \rho}{d \ln t} = \rho^{*}[\rho] - \rho
\end{equation}
Letting $\alpha = \ln t$,
\begin{equation}
\frac{d \rho}{d \alpha} = \rho^{*}[\rho] - \rho
\label{eq9}
\end{equation}
We note that
for $\rho < \rho^* $, $\rho$  increases,
whereas for  $\rho > \rho^* $, $\rho$ decreases.
Therefore $\rho = \rho^*$ informally appears to be a stable fixed point.

We now consider more carefully the function $\rho^*[\rho]$.
Letting $t \gg M$, we find
\begin{equation}
\rho (t) = \frac{M}{t} ({\rm arbitrary\: initial} ~ \rho)
 + ( 1 - \frac{M}{t}) \sum^{t/m}_{i=M/m} 
 \frac{\rho_i}{(t-M)/m}
\label{eq10}
\end{equation}
where the density at
time $t = i  \Delta t$, $\rho_i = \rho^{*}_i = e^{-U} / P_i (\lambda)$,
is correct for a given $\lambda$,
$\rho_i({\bf x}) / \rho_i({\bf y}) =
e^{-\beta U({\bf x}) + \beta U({\bf y})}$,
 but for which $P_i(\lambda)$ 
has not converged to Eqn.\ \ref{eq0}.
This result for the ratio of the density follows from
Eqns.\ \ref{eq2}--\ref{eq4}.
Thus, for a given $\lambda$,
\begin{equation}
\rho(t)({\bf x} ) = ({\rm const})   \exp^{-U({\bf x})} + O(\frac{M}{t}) \ .
\end{equation}
Eqns.\ \ref{eq2}--\ref{eq4} 
in the limit $1 \ll m^2 \ll M \ll t$ thus imply
that $\hat F$ and $\rho$ are becoming exact:
\begin{eqnarray}
\rho^{*}[\rho] & = & 
\rho^* + \frac{\delta \rho^*}{\delta \rho} \delta \rho\nonumber \\
& = &
\rho^* + O(M/t) \ .
\end{eqnarray}
Thus, Eqn.\ \ref{eq9} becomes
\begin{eqnarray}
\frac{d \rho}{d \alpha} & = &\rho^{*}[\rho^* + \delta \rho] - \rho +
 O(M/t, m^2/t, r^m, 1/\sqrt m)\nonumber \\
& = &
\rho^{*} - \rho + O(M e^{- \alpha}, m^2 e^{-\alpha}, r^m, 1/\sqrt m  ) \ ,
\end{eqnarray}
where we have reinserted the additional error terms from Eqn.\ \ref{eq7}.
Thus, 
\begin{equation}
\rho \rightarrow \rho^* + O(M e^{-\alpha}, m^2 e^{-\alpha}, r^m, 1/\sqrt m  )
\end{equation}
exponentially fast in $\alpha$.
Thus, in the limit $ 1 \ll m^2 \ll M \ll t$,
we find
\begin{equation}
\rho(t) \rightarrow \rho^* + O(M/t, m^2/t, r^m, 1/\sqrt m, 1/t )
\label{eqnfin}
\end{equation}
as $t \rightarrow \infty$, 
where the errors in
Eqn.\ \ref{eqnfin} 
are respectively from the error on $\hat F$, 
the change of the density during $\Delta t$, 
the Monte Carlo convergence for a given $\hat F$, 
the stochastic convergence of the density to the distribution, and 
the convergence of the differential equation.

\section{Discussion}
\label{sec:discussion}
Our analysis of the convergence of the adaptive integration Monte Carlo 
scheme gives insight into why the convergence of the method is so rapid.
That is, the adaptive method converges as fast as the underlying Monte Carlo
method converges for density distributions with a given
value of $\lambda$.  Once these estimates have converged, then the sampling
over different values of $\lambda$ is typically exponentially accelerated
by the importance sampling bias introduced by the $P(\lambda)$
factors in Eqn.\ \ref{eq2}, so that a simple random walk in the
$\lambda$ order parameter may be achieved without any barriers to sampling.
Thus, if the most significant barriers in the problem are fully
captured by the $\lambda$ order parameter, then the adaptive
integration dramatically speeds up the convergence of the underlying
Monte Carlo algorithm.  In the context of
the above analysis, an improved importance sampling
estimate dramatically reduces the number of steps, $m$, that it
takes to reach equilibrium for a given value of $\lambda$.

In the above analysis, full convergence was shown.  That is, convergence
of both the density distribution for a given value of $\lambda$
and the distribution of $\lambda$ was shown.  Since adaptive
integration Monte Carlo is a form of iteration,
we see the linear convergence, as $O(1/t)$, in Eqn.\ \ref{eqnfin}.
In the analysis of the Monte Carlo data, we will remove the
bias introduced by the $P(\lambda)$.  That is, we will adjust
the observed densities, scaling out the $P(\lambda)$ dependence
by histogram reweighting.
Incidentally, we note from Eqn.\ \ref{eq10}, that when reweighting the 
histograms, one should use the the average of the calculated
importance sampling factors, $\langle 1/P_i(\lambda) \rangle$, rather than
the instantaneous importance sampling factor, $1/P_t(\lambda)$.
Such a reweighting procedure implies by Eqns.\ \ref{eq2}--\ref{eq3}
that once the simulated density has converged
within a given value of $\lambda$ (in $O(m)$ steps), 
the reweighted density has also converged for all $\lambda$.
So, the slow linear convergence observed in Eqn.\ \ref{eqnfin}
is actually not a great concern.  Conversely, our major algorithmic interest
is in the exponential reduction of the sampling time, $m$, within
a given value of $\lambda$, which is already induced by
an only roughly converged importance sampling bias, $P(\lambda)$.

In conclusion, detailed balance, or balance,\cite{Deem_balance} need
not be satisfied at all times in a Monte Carlo simulation.  Balance
need only be satisfied asymptotically.  Indeed, the desire to
maintain balance comes only from the Markov chain theory that
shows such an approach converges to a unique limiting distribution.
Any Monte Carlo
method that converges to the specified unique limiting distribution
will be equally valid.  
Given the success of the adaptive Monte Carlo methods,
it would appear that the importance
of detailed balance is over-rated.

\section*{Acknowledgments}
The authors thank David J.\ Aldous for 
pointing out the relevance of the contraction condition, Eqn.\ \ref{aldous}.
This work was supported by the U.S.\ Department of Energy Office of Basic
Energy Sciences.

\bibliography{chandler}

\end{document}